\documentclass[%
 aip,
 amsmath,amssymb,
 reprint,%
]{revtex4-1}

\usepackage{graphicx}
\usepackage{dcolumn}
\usepackage{bm}

\usepackage[utf8]{inputenc}
\usepackage[T1]{fontenc}
\usepackage{mathptmx}
\usepackage{etoolbox}

\makeatletter
\def\@email#1#2{%
 \endgroup
 \patchcmd{\titleblock@produce}
  {\frontmatter@RRAPformat}
  {\frontmatter@RRAPformat{\produce@RRAP{*#1\href{mailto:#2}{#2}}}\frontmatter@RRAPformat}
  {}{}
}%
\makeatother
\begin{document}

\preprint{AIP/123-QED}

\title[Purely optical macroscopic trap for alkaline-earth and similar atoms]{Purely optical macroscopic trap for alkaline-earth and similar atoms}
\author{O.N. Prudnikov}
\email{oleg.nsu@gmail.com}
\affiliation{Institute of Laser Physics, Laser Physics Department, 630090, Novosibirsk, Russia}
\affiliation{Novosibirsk State University, Physics Department, 630090, Novosibirsk, Russia}

\author{V. I. Yudin}
\affiliation{Institute of Laser Physics, Laser Physics Department, 630090, Novosibirsk, Russia}
\affiliation{Novosibirsk State University, Physics Department, 630090, Novosibirsk, Russia}
\affiliation{Novosibirsk State Technical University, Physics and Technology Department, 630073, Novosibirsk, Russia}

\author{R. Ya. Ilenkov}
\affiliation{Institute of Laser Physics, Laser Physics Department, 630090, Novosibirsk, Russia}

\author{A. V. Taichenachev}
\affiliation{Institute of Laser Physics, Laser Physics Department, 630090, Novosibirsk, Russia}
\affiliation{Novosibirsk State University, Physics Department, 630090, Novosibirsk, Russia}

\date{\today}

\begin{abstract}
We consider a laser cooling and trapping of alkaline-earth and similar atoms in a bichromatic field resonant to a closed optical transition $^1S_0 \to \, ^1P_1$ or $^1S_0 \to \, ^3P_1$. It is shown that new kinetic effects emerge compared to monochromatic fields, enabling the formation of a deep macroscopic trap capable of capturing and cooling neutral atoms to sub-Doppler temperatures. Such a purely optical macroscopic trap can serve as an alternative to the well-known magneto-optical trap and can be used in applications requiring minimization of the magnetic field in the cold atom cloud region. The obtained results are of interest for the new generation of quantum sensors and optical frequency standards.
\end{abstract}


\maketitle

\section{\label{sec:level1}Introduction}
As is well known \cite{kaz}, efficient laser cooling of atoms is possible due to the action of Doppler and sub-Doppler mechanisms leading to dissipation of the atom kinetic  energy in the field of counter-propagating waves resonant to the optical transition. However, in most cases, experiments with cold atoms requires not only cooling to low temperatures, 
but also reliable trapping. This has stimulated work on searching for configurations of light fields capable of forming deep dissipative traps for cooling and holding of atoms. The most successful solution to this problem is the magneto-optical trap (MOT), proposed and experimentally realized in the late 1980s \cite{raab1987}. The unique combination of an inhomogeneous magnetic field and counter-propagating light waves forms a deep magneto-optical potential, allowing for cooling and trapping of atoms. 

On the other hand, the creation of modern high-precision quantum devices based on cold atoms (such as atomic clocks, quantum logic gates, quantum memory, atomic interferometers, gravimeters, gyroscopes, etc.) imposes new requirements that may pose significant challenges when using a standard magneto-optical trap. One such requirement is the precise control of the magnetic field in the region where cold atoms are interrogated as a quantum object. Therefore, the search for new mechanisms for cooling and trapping atoms without use of magnetic field is particularly relevant, especially in compact devices. For example in Ref.\, \cite{pru2023}, we proposed the idea of deep dissipative potential of macroscopic scale in bichromatic field for cooling and trapping of Li atoms, which have a specific structure of energy levels. The  components of the bichromatic field are resonant to different optical transitions, allowing the formation of purely optical macroscopic traps with relatively low field intensities comparable to those used in MOTs. The formed deep dissipative trap is determined by a combination of two effects. First, the dipole force rectification in a bichromatic field, predicted in \cite{kaz87,ovch90} and used for effective slowing of atomic beams \cite{soding,Chieda,narong}. Second, dissipative mechanisms of laser cooling  in a bichromatic field \cite{pru2013,pru2017}. The spatial beating of optical shifts of atomic levels in a bichromatic field forms a macroscopic potential with a period $\Lambda = \pi/|\Delta {\bf k}|$, where $\Delta {\bf k}={\bf k}_2-{\bf k}_1$ is the difference in the wave vectors of the frequency components of the bichromatic field. When the frequency difference is chosen in the range of $5 - 30$ GHz, the macroscopic period of the optical trap has a centimeter scale, enabling sub-millimeter sizes of the trapped atoms cloud. For demonstration of this effect, in \cite{pru2023}, transitions between fine or hyperfine structure levels of lithium atoms were considered. However, not for all elements the optical transitions with the above frequency differences can be found.

In the present work, we are developing the idea of purely optical macroscopic dissipative trap for the atoms in which the energy levels structure does not contain transitions with frequency differences in the range of $5 - 30$ GHz. For these atoms to create a deep macroscopic potential, we consider a bichromatic field with the required frequency difference interacting with the same atomic transition. The case under consideration does not require a specific structure of fine or hyperfine levels (as in \cite{pru2023}) and can be realized for most alkaline-earth-like atoms (e.g., Ca, Sr, Ba, Hg, Mg, Yb, see Table \ref{optical_transitions}), which have a dipole-allowed closed optical transition. These elements are of particular interest for optical frequency standards \cite{Falke,Poli2013,ludlow,kulosa,Goncharov,Katori2020}, quantum memory \cite{Covey}, and atomic interferometers \cite{taich,Riehle,Rudolph,Hu}. We demonstrate that a deep dissipative macroscopic trapping potential can be formed. Moreover, sub-Doppler temperature can be reached in it for odd isotopes.

\begin{table*}[tbp]
\begin{center}
\begin{tabular}{ccccccccccccc}
\hline \hline \\
element  && nuclear&& laser cooling        &&  $\gamma/2\pi$&&$I_S$  &&$\lambda$&& $\varepsilon_R$\\
         && spin &&transition   &&   (MHz)  &&(mW/cm$^2$)&& (nm) &&             \\ \hline \\

$^{25}$Mg,\, $^{24}$Mg && 5/2,\, 0  &&$3^1$S$_{0} \to 3^{1}$P$_{1}$&& 79 && 439 &&  285.3 &&  $1.3\cdot 10^{-3}$        \\\\
$^{40}$Ca && 0  &&$4^1$S$_{0} \to 4^1$P$_{1}$&& 35.7 && 61.9 && 422.6 &&
$7.8\cdot 10^{-4}$
\\\\
$^{87}$Sr\,, $^{88}$Sr && 9/2,\, 0  &&$5^1$S$_{0} \to 5^1$P$_{1}$&& 32 && 42.7 &&  461 &&  $3.4\cdot 10^{-4}$ \\
            &&   &&$5^1$S$_{0} \to 5^3$P$_{1}$&& $7.5\cdot 10^{-3}$&& $2.8\cdot10^{-3}$ && 689&& 0.64\\\\
$^{135}$Ba,\, $^{138}$Ba && 3/2,\, 0  &&$6^1$S$_{0} \to 6^1$P$_{1}$&& 18.9 && 14.6 && 553.7 &&
$2.5\cdot 10^{-4}$
\\
            &&   &&$6^1$S$_{0} \to 6^3$P$_{1}$&& $0.048$&& $12.5\cdot 10^{-3}$ && 791.3&& 0.048\\\\
$^{171}$Yb,\,$^{174}$Yb && 1/2,\, 0  &&$6^1$S$_{0} \to 6^1$P$_{1}$&& 29 &&59  &&  399 &&  $2.5\cdot 10^{-4}$ \\
            &&   &&$6^1$S$_{0} \to 6^3$P$_{1}$&& 0.18 &&0.14  &&  555.8 &&  0.02\\\\

$^{199}$Hg,\, $^{200}$Hg && 1/2,\, 0  &&$6^1$S$_{0} \to 6^1$P$_{1}$&& 120 &&2480 &&  185 &&  $2.4\cdot 10^{-4}$ \\
            &&   &&$6^1$S$_{0} \to 6^3$P$_{1}$&& 1.3 &&10.6  &&  253.7 &&  0.012\\\\
\hline\hline
\end{tabular}
\end{center} \caption{Optical transitions of alkaline-earth-like atoms for laser cooling, their parameters (data from \cite{taich, ludlow, adams, baatoms1, baatoms2, Ybatoms}): wavelength $\lambda$,
saturation intensity ($I_S=2\pi^2 \hbar c \gamma/\lambda^3$) and recoil parameter $\varepsilon_R = \hbar k^2/(2M\gamma)$ is equal to the ratio of the recoil energy to the natural linewidth (M is the mass of the atom).
}\label{optical_transitions}
\end{table*}

\section{Kinetics of atoms in a bichromatic field}
Let us consider the interaction of atoms with a bichromatic laser field having frequency components $\omega_1$ and $\omega_2$
\begin{equation}\label{field}
    {\bf E}({\bf r},t) = \mbox{Re}\Bigg\{\sum_{\alpha=1,2} {\bf E}_{\alpha}({\bf r}) e^{-i\,\omega_\alpha t} \Bigg\}\,,
\end{equation}
which are resonant to an optical transition with frequency $\omega_0$. The optical transitions used for laser cooling of alkaline-earth-like atoms have the electron configuration ${\nobreak J_g=0 \to J_e=1}$ (see Table \ref{optical_transitions}), where $J_g$ and $J_e$ are the electron angular momenta of the ground $(g)$ and excited $(e)$ states, respectively. For odd isotopes with non-zero nuclear spin $I$ (see Table \ref{optical_transitions}), a set of excited states emerges with total angular momentum $F_{e_n}$, where $|J_e-I| \leq F_{e_n} \leq |J_e+I|$, while the ground state has angular momentum $F_g=I$. Such level structure exists for the dipole transition $^1S_0 \to \,^1P_1$ and  intercombination dipole transition $^1S_0 \to \,^3P_1$. It should be noted that the natural linewidth of the intercombination transition increases with the atomic number, which allows it to be used for laser cooling of heavy elements \cite{Maruyama_Yb,Katori_Sr,intercomb}.

The Hamiltonian of an atom interacting with a laser field has the form
\begin{equation}
{\hat H} = \frac{{\hat p}^2}{2M} + {\hat H}_0 +{\hat W}(t)\,,
\end{equation}
where the first term is the kinetic energy operator, the second is the Hamiltonian of the atom in the rest frame, and the last term is atom-light interaction part. The Hamiltonian ${\hat H}_0$ can be written in a form ${\hat H}_0 = {\hat H}_0^{ee}+{\hat H}_0^{gg}$, where the blocks for the ground and excited states are defined through the projection operators ${\hat P}^{e_n}$ and ${\hat P}^{g}$
as follows
\begin{eqnarray}
{\hat H}_0^{ee}&=&  \sum_{n} {\cal E}_{e_n} {\hat P}^{e_n} \,,\,\,\,  {\hat P}^{e_n} = \sum_{M_{e_n}} |F_{e_n}, M_{e_n} \rangle \langle F_{e_n}, M_{e_n} |      \nonumber \\
{\hat H}_0^{gg}&=&  {\cal E}_g{\hat P}^{g}\,,\,\,\, {\hat P}^{g} =\sum_{M_{g}}|F_{g}, M_{g} \rangle \langle F_{g}, M_{g} | \, .
\end{eqnarray}
Here ${\cal E}_{e_n}$ is the energy of the $n$-th excited state, ${\cal E}_{g}$ is the energy of the ground state,  $|F_{e_n}, M_{e_n} \rangle$ and $|F_{g}, M_{g} \rangle$ describe the wave functions of the Zeeman sublevels in the excited and ground states, where $M_{e_n}$ and $M_{g}$ are magnetic quantum numbers ($-F_{e_n}\leq M_{e_n}\leq F_{e_n}$, $-F_{g}\leq M_{g}\leq F_{g}$). The atom-light interaction in the electric dipole approximation can be written as
\begin{eqnarray}
{\hat W}({\bf r},t) &=& \hbar \sum_{\alpha=1,2}{\hat V}_\alpha({\bf r}) \, e^{-i\omega_\alpha t} +h.c. \, , \nonumber \\
{\hat V}_\alpha({\bf r}) &=& -\left( {\bf E}_\alpha({\bf r}) \cdot {\hat {\bf D}}\right){\overline d}/2\hbar\, ,
\end{eqnarray}
where ${\overline d}$ is the reduced matrix element of the dipole transition for considered electron transition $J_g=0 \to J_e=1$, ${\hat {\bf D}}$ is the normalized (dimensionless) operator of the dipole moment of the atom, whose components ${\hat D}_s$  ($s=0,\pm1$) in the cyclic basis ${\nobreak \{{\bf e}_{\pm} = \mp ({\bf e}_x \pm i {\bf e}_y)/\sqrt{2}, {\bf e}_0= {\bf e}_z \}}$ are expressed through the Clebsch-Gordan coefficients
\begin{equation}
{\hat D}_s = \sum_{F_{e_n}, M_{e_n},M_g}  | F_{e_n}, M_{e_n} \rangle \, C^{F_{e_n},M_{e_n}}_{F_g, M_g; 1,s} \,\langle F_g, M_g| \, .
\end{equation}
The quantum kinetic equation describing the evolution of the atomic density matrix in the field (\ref{field}) with full consideration of recoil effects has the following form in coordinate representation
\begin{eqnarray}\label{GDE}
&&\frac{\partial}{\partial t}{\hat \rho}({\bf r}_1,{\bf r}_2) =  \\
&&-\frac{i}{\hbar} \left[{\hat H}({\bf r}_1){\hat \rho}({\bf r}_1,{\bf r}_2)- {\hat \rho}({\bf r}_1,{\bf r}_2){\hat H}({\bf r}_2)\right] -{\hat \Gamma}\{{\hat \rho}({\bf r}_1,{\bf r}_2)\} \, . \nonumber 
\end{eqnarray}
The last term describes the non-Hamiltonian evolution of the quantum system as a result of spontaneous relaxation taking into account recoil effects (see, for example, \cite{prud2011Stationary}):
\begin{eqnarray}\label{e:GE1}
&\hat \Gamma \{ \hat \rho \}& = \frac{\gamma}{2} (\hat P^e \hat \rho + \hat \rho \hat P^e)
- \hat{\gamma}\left\{
\hat{\rho} \right\}  \\
&\hat{\gamma}\left\{
\hat{\rho} \right\}&= \gamma \frac{3}{2} \Big \langle \sum_{\xi=1,2} (\hat {\bf D} \cdot {\bf e}_\xi (\mathbf{k}))^\dagger e^{-i \mathbf{k} \hat{\mathbf{r}}_1} \hat \rho
e^{i \mathbf{k} \hat {\mathbf{r}}_2}(\hat {\bf D} \cdot {\bf e}_\xi (\mathbf{k})) \Big\rangle\,. \nonumber
\end{eqnarray}
Here ${\hat P}^e = \sum_{n} {\hat P}^{e_n}$ is the full projector onto all excited states, and $\gamma$ is the natural linewidth. The operator $\hat{\gamma}\left\{ \hat{\rho} \right\}$ describes the arrival to the ground state from excited states due to spontaneous emission, where $\langle \ldots \rangle$ means averaging over the emission directions of spontaneous photons with momentum $\hbar k$ and two mutually orthogonal polarizations ${\bf e}_\xi (\mathbf{k})$.

As noted in the introduction, to create a deep dissipative potential with a centimeter scale, it is necessary to use the fields with a frequency difference of $|\omega_1-\omega_2|\sim 5-30\,$ GHz.
Such a difference in frequencies allows us to neglect the rapidly oscillating contributions $\mbox{exp}(\pm i (\omega_1 - \omega_2)t$) in equation (\ref{GDE}). As a result, in the rotating-wave approximation, we obtain the following system of equations for the matrix blocks $\hat{\rho}({\bf r}_1,{\bf r}_2)$
\begin{widetext} 
\begin{eqnarray}\label{xx}
\label{block} \frac{d}{dt} \hat{\rho}^{ee}({\bf r}_1,{\bf r}_2) &=& -\gamma
\hat{\rho}^{ee}({\bf r}_1,{\bf r}_2) - \frac{i}{\hbar} \Big[{\hat H}_0^{ee}{\hat \rho}^{ee}({\bf r}_1,{\bf r}_2)-{\hat \rho}^{ee}({\bf r}_1,{\bf r}_2){\hat H}_0^{ee}\Big]\nonumber \\
&&-i\sum_{\alpha = 1,2}\left( \hat{V}_\alpha({\bf r}_1)
\hat{\rho}_\alpha^{ge}({\bf r}_1,{\bf r}_2)-\hat{\rho}_\alpha^{eg}({\bf r}_1,{\bf r}_2)\hat{V}^{\dagger}_\alpha({\bf r}_2)
\right)\nonumber \\
\frac{d}{dt} \hat{\rho}^{gg}({\bf r}_1,{\bf r}_2) &=& \hat{\gamma}\left\{
\hat{\rho}^{ee}({\bf r}_1,{\bf r}_2) \right\} - \frac{i}{\hbar} \Big[{\hat H}_0^{gg}{\hat \rho}^{gg}({\bf r}_1,{\bf r}_2)-{\hat \rho}^{gg}({\bf r}_1,{\bf r}_2){\hat H}_0^{gg}\Big]\nonumber \\
&&-i\sum_{\alpha=1,2}\left( \hat{V}_\alpha^{\dagger}({\bf r}_1)
\hat{\rho}_\alpha^{eg}({\bf r}_1,{\bf r}_2) -\hat{\rho}_\alpha^{ge}({\bf r}_1,{\bf r}_2)\hat{V}_\alpha({\bf r}_2)
\right) \nonumber \\
\frac{d}{dt} \hat{\rho}^{eg}_\alpha({\bf r}_1,{\bf r}_2) &=&
-\left(\frac{\gamma}{2}-i\omega_{\alpha} \right) \hat{\rho}^{eg}_\alpha({\bf r}_1,{\bf r}_2)
-\frac{i}{\hbar} \Big[{\hat H}_0^{ee}{\hat \rho}^{eg}({\bf r}_1,{\bf r}_2)-{\hat \rho}^{eg}({\bf r}_1,{\bf r}_2){\hat H}_0^{gg}\Big]
\nonumber\\
&&-i\left(
\hat{V}_\alpha({\bf r}_1) \hat{\rho}^{gg}({\bf r}_1,{\bf r}_2)
-\hat{\rho}^{ee}({\bf r}_1,{\bf r}_2)\hat{V}_\alpha({\bf r}_2) \right),\;\;\alpha=1,2 \nonumber \\
\frac{d}{dt} \hat{\rho}^{ge}_\alpha({\bf r}_1,{\bf r}_2)&=&
-\left(\frac{\gamma}{2}+i\omega_\alpha\right)\hat{\rho}^{ge}_1({\bf r}_1,{\bf r}_2) -\frac{i}{\hbar}\Big[{\hat H}_0^{gg}{\hat \rho}^{ge}({\bf r}_1,{\bf r}_2)-{\hat \rho}^{ge}({\bf r}_1,{\bf r}_2){\hat H}_0^{ee}\Big]\nonumber\\
&& -i\left(
\hat{V}_\alpha^{\dagger}({\bf r}_1)\hat{\rho}^{ee}({\bf r}_1,{\bf r}_2)
-\hat{\rho}^{gg}({\bf r}_1,{\bf r}_2)\hat{V}_\alpha^{\dagger}({\bf r}_2) \right) ,\;\;\alpha=1,2 \, ,
\end{eqnarray}
\end{widetext}
where matrix blocks are defined by the following
\begin{eqnarray}
&&\hat{\rho}^{gg} = {\hat P}^{g} {\hat \rho} {\hat P}^{g}\,,\;\; \hat{\rho}^{ee} = {\hat P}^{e} {\hat \rho} {\hat P}^{e}\,, \\
&&\hat{\rho}^{eg} = {\hat P}^{e} {\hat \rho} {\hat P}^{g}=\hat{\rho}^{eg}_1
\mbox{e}^{-i\omega_1 t}+\hat{\rho}^{eg}_2 \mbox{e}^{-i\omega_2 t} \, ,\;\;\; \hat{\rho}^{ge} = \big(\hat{\rho}^{eg}\big)^{\dagger}\,, \nonumber 
\end{eqnarray}
Here $\hat{\rho}^{gg}$ and $\hat{\rho}^{ee}$ describe atoms in the ground and excited states, $\hat{\rho}^{eg}_{\alpha}$, $\hat{\rho}^{ge}_{\alpha}$ ($\alpha = 1,2$) are the slow amplitudes of the off-diagonal elements of the density matrix.
For brevity in (\ref{xx}), we use 
\begin{equation}
\frac{d}{dt} =
\frac{\partial}{\partial t} -i\frac{\hbar}{2M}\, \frac{\partial^2}{\partial{\bf r}_1^2} +i\frac{\hbar}{2M}\, \frac{\partial^2}{\partial{\bf r}_2^2}\,.
\end{equation}
As well known \cite{Dalibard1985,Javanainen1991,Javanainen1992}, the equation for the atomic density matrix in the Wigner representation $\hat{\rho}({\bf r},{\bf p})$ by means of expansion in terms of the recoil parameter $\hbar k/\Delta p\ll 1$ ($\Delta p$ is the width of the momentum distribution of the ensemble of atoms) can be reduced to the Fokker-Planck equation for the distribution function in the phase space ${\cal F}({\bf r},{\bf p}) =
\mbox{Tr}\{\hat{\rho}({\bf r},{\bf p})\}$ 
\begin{eqnarray}\label{FP}
 \Bigg(\frac{\partial}{\partial t}
&+&\frac{{\bf p}}{M}\cdot \frac{\partial}{\partial {\bf r}} \Bigg)\,{\cal F}({\bf r},{\bf p})  =-
\sum_j \frac{\partial}{\partial {p_j}}F_j({\bf r},{\bf p}) \,{\cal F}({\bf r},{\bf p})  \nonumber \\
&+&\sum_{jj'} \frac{\partial}{\partial {p_j}}\frac{\partial}{\partial {p_{j'}}} D_{jj'}({\bf r},{\bf p}) \,{\cal F}({\bf r},{\bf p}) \, ,
\end{eqnarray}
with coefficients corresponding
to the force ${\bf F}({\bf r},{\bf p})$ acting on the atom and the diffusion tensor $D_{jj'}({\bf r},{\bf p})$. 
Note that the results obtained within the semiclassical approach based on the Fokker-Planck equation (\ref{FP}) are well agreed with the results of direct quantum simulation in the case of  small recoil parameter $\varepsilon_R = \hbar k^2/(2M\gamma) \ll 1$ \cite{pru2007,kirp2020,kirp2022}. Since the condition $\varepsilon_R \ll 1$ is satisfied quite well for the elements under consideration (see table \ref{optical_transitions}), below we will use semiclassical approximation.

To obtain expressions for the force and diffusion, we use the method proposed in \cite{prudnikov2003,prudnikov2004}. The essence of the method is that the  expansion of the quantum kinetic equation on recoil parameter $\hbar k/\Delta p\ll 1$ in coordinate representation (\ref{GDE}),
 rewritten in variables
\begin{equation}\label{qq}
{\bf q} = {\bf r}_1-{\bf r}_2 \,,\;\;\;
{\bf r} = ({\bf r}_1+{\bf r}_2)/2\,,
\end{equation}
is equivalent to expansion in powers of $-ikq_j$, where $q_j$ are the Cartesian components of the vector ${\bf q}$. This expansion can be written in the general operator form
\begin{eqnarray}\label{razlozenie}
  \bigg(\frac{\partial}{\partial t} &-&i\frac{\hbar}{M}\frac{\partial}{\partial {\bf q}} \cdot \frac{\partial}{\partial {\bf r}}\bigg)\hat{\rho}({\bf r},{\bf q}) = \hat{{\cal
L}}^{(0)}\{\hat{\rho} \}  \\
&-&\sum_j ikq_j\, \hat{{\cal L}}^{(1)}_j\{\hat{\rho}
  \}+ \sum_{jj'}(-ik)^2q_jq_{j'} \hat{{\cal
L}}^{(2)}_{jj'}\{\hat{\rho} \} \dots \, , \nonumber
\end{eqnarray}
where the explicit expressions of the operators $\hat{{\cal L}}^{(k)}$ are given in Appendix {\bf A}. Note that the transition from coordinate representation (\ref{qq}) to Wigner is defined by Fourier transformation
\begin{equation}
    \hat{\rho}({\bf r},{\bf p}) = \int \hat{\rho}({\bf r},{\bf q}) e^{-i {\bf p} {\bf q}/\hbar} d{\bf q}/(2\pi \hbar)^3 \,.
\end{equation}
Following the proposed  methods \cite{prudnikov2003,prudnikov2004}, the equation
(\ref{razlozenie}) within the semiclassical approach can be reduced to the Fokker-Planck equation (\ref{FP}) for the distribution function in the phase space ${\cal F}({\bf r},{\bf p})$. In this case, the expression for the force ${\bf F}({\bf r},{\bf p})$
is determined by the first-order terms in the expansion
(\ref{razlozenie}). The Cartesian component of the force $F_j$ acting on the atom with the velocity ${\bf v}={\bf p}/M$ has the form:
\begin{equation}
\label{force_origin}
F_j({\bf r},{\bf p}) = - \mbox{Tr}\Big\{ \hat{{\cal
L}}^{(1)}_j\{\hat{\sigma}({\bf r},{\bf p})\} \Big\} \, ,
\end{equation}
where $\hat{\sigma}({\bf r},{\bf p})$ is a stationary solution of the optical Bloch equation
\begin{equation}
\frac{{\bf p}}{M}\cdot \frac{\partial}{\partial {\bf r}}\,\hat{\sigma} = \hat{{\cal L}}^{(0)}\{\hat{\sigma} \} \, , \;\;\mbox{Tr}\big\{ \hat{\sigma} \big\}=1 \,.
\end{equation}
The diffusion coefficient is split into the sum of two contributions
\begin{equation}\label{diffusion}
D_{jj'}=D_{jj'}^{(sp)}+D_{jj'}^{(ind)} \, .
\end{equation}
The first term in it is the spontaneous diffusion tensor, determined by the fluctuation of the atomic momentum during spontaneous emission of photons
\begin{eqnarray}\label{dsp}
&&D_{jj'}^{(sp)}({\bf r},{\bf p}) = \hbar^2 k^2 \mbox{Tr}\Big\{ \hat{{\cal
L}}^{(2)}_{jj'}\{\hat{\sigma}\} \Big\}=  \\
&&\frac{\hbar^2k^2\gamma}{5}\mbox{Tr}\Bigg\{
\Big( \delta_{jj'}{\hat P}^e -\frac{1}{4}\big({\hat D}_j{\hat D}_{j'}^{\dagger} +{\hat D}_{j'}{\hat D}_{j}^{\dagger} \big)\Big)
{\hat \sigma}({\bf r},{\bf p}) \Bigg\}\,, \nonumber
\end{eqnarray}
where $\delta_{jj'}$ is the Kronecker symbol. The second term in (\ref{diffusion})
is the tensor of induced diffusion, is determined by the fluctuation of the atomic momentum in the processes of induced absorption/emission of field photons
\begin{equation}\label{dind}
D_{jj'}^{(ind)} = \hbar^2 k^2 \mbox{Tr}\Big\{ \hat{{\cal
L}}^{(1)}_{j}\{\hat{\eta}_{j'}\} \Big\} \, ,
\end{equation}
where the matrix $\hat{\eta}_{j}({\bf r},{\bf p})$ satisfies the equation
\begin{eqnarray}\label{eta}
\frac{{\bf p}}{M}\cdot \frac{\partial}{\partial {\bf r}}\,\hat{\eta}_{j} &=& \hat{{\cal L}}^{(0)}\{\hat{\eta}_{j} \} + \hat{{\cal L}}^{(1)}_{j}\{\hat{\sigma} \} -{\hat{\sigma}}\,\mbox{Tr}\Big\{ \hat{{\cal
L}}^{(1)}_j\{\hat{\sigma}\} \Big\}\nonumber \\
\mbox{Tr}\big\{ \hat{\eta}_{j} \big\}&=&0\,,
\end{eqnarray}
which is equivalent to the approach developed in \cite{Javanainen1991,Javanainen1992}.

Note that for the model of a two-level atom  \cite{pru2013,pru2017} we obtained analytical expressions for the optical potential, friction coefficients (linear in velocity correction to the force) and the diffusion tensor in the limit of low velocities. However, to analyze the possibility of sub-Doppler laser cooling it is necessary to get expressions for the force and diffusion beyond the low-velocity approximation $kv\ll \gamma$. The equations (\ref{force_origin})-(\ref{eta}) allow us to obtain the required expressions for the force and diffusion tensors in a bichromatic field and to analyze sub-Doppler cooling taking into account the real hyperfine and Zeeman structure of atomic levels.

\begin{figure}[t]
    \centering
    \centerline{\includegraphics[width=3 in]{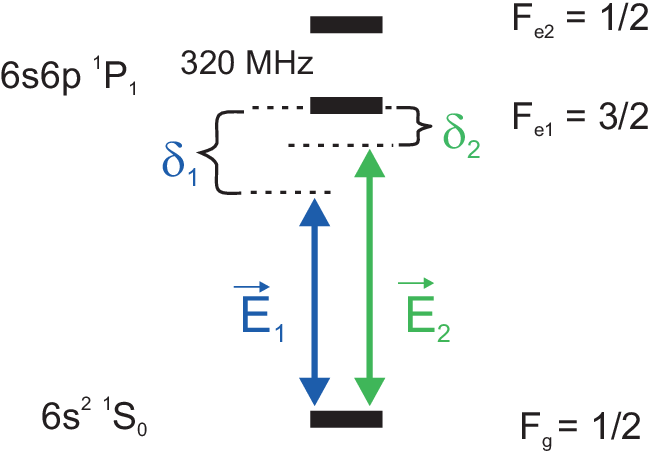}}
    \caption{The scheme of energy levels of ytterbium-171 for laser cooling and trapping in a bichromatic field. Solid arrows represents transitions induced  by components of the bichromatic field ${\bf E}_1$ and ${\bf E}_2$.}
    \label{fig:F1}
\end{figure}

\begin{figure}[b]
\centerline{\includegraphics[width=3.2 in]{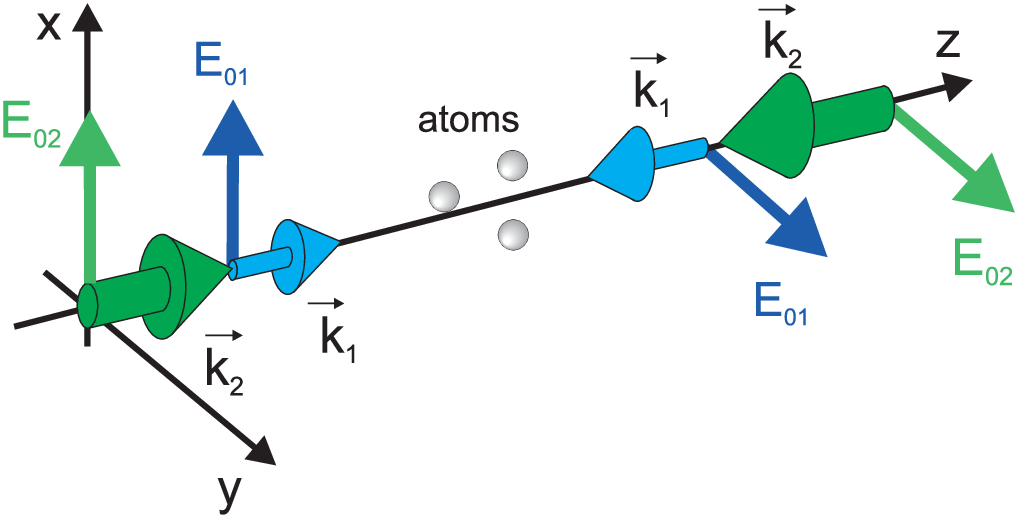}}
\caption{Double $lin \perp lin$ bichromatic field configuration.} \label{fig:F2}
\end{figure}

\begin{figure}[t]
\centerline{\includegraphics[width=3.0 in]{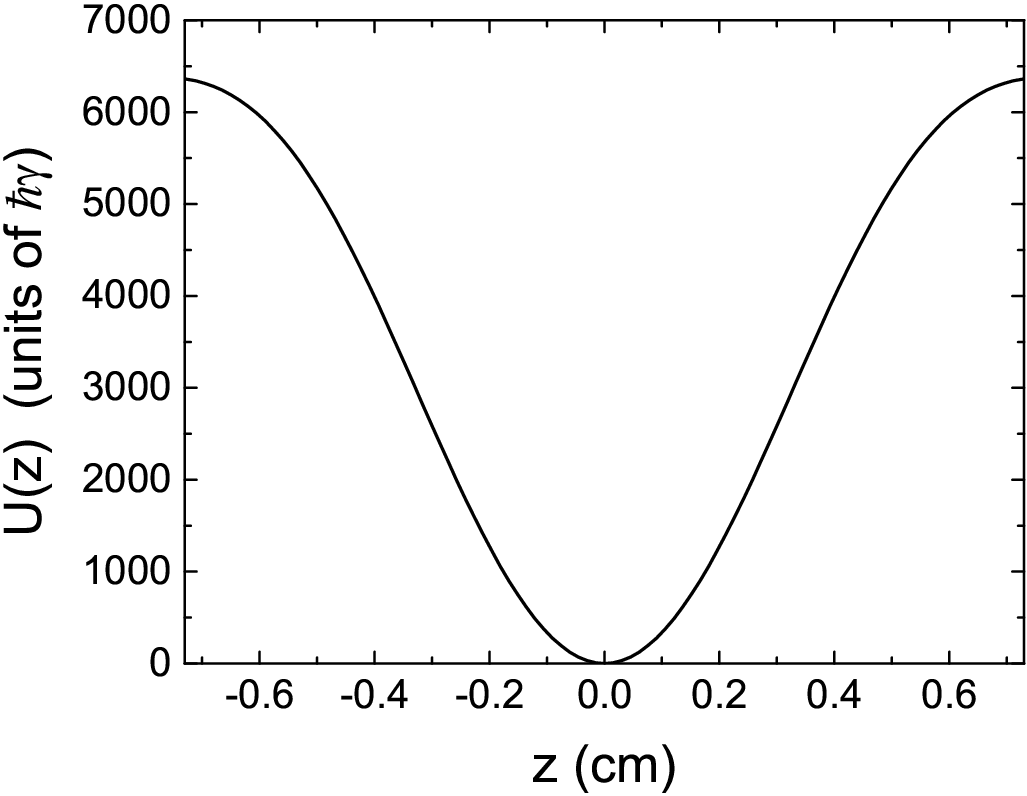}}
\caption{Macroscopic optical potential $U(z)$ in units of $\hbar \gamma$ in a bichromatic field. Field detunings $\delta_1 = -346 \gamma$ ($\simeq 10\,$GHz), and $\delta_2 = -3 \gamma$. Intensities of the waves of the frequency components $I_1 \simeq 57$ W/cm$^2$ ($\Omega_{01} \simeq 22 \gamma$), and $I_2 \simeq 0.4$ W/cm$^2$ ($\Omega_{02} \simeq 1.9 \gamma$).} \label{fig:F3}
\end{figure}

\section{Numerical calculations of laser cooling and trapping of atoms}
As a specific example, in this section we consider the $^{171}$Yb atom, whose levels are degenerate in the projection of the angular momentum (Fig.\,\ref{fig:F1}), which is a necessary condition for sub-Doppler cooling \cite{dalibard89}. In this case, we consider the double $lin \perp lin$ configuration of the light field proposed by us in \cite{pru2023} to form a macroscopic potential. In this case, both fields ${\bf E}_1$ and ${\bf E}_2$ in the expression (\ref{field}) are formed by counterpropagating waves with orthogonal linear polarizations (Fig.\,\ref{fig:F2})
\begin{equation}\label{double_linlin}
      {\bf E}_\alpha(z)=E_{0\alpha} \Big({\bf e}_x e^{ik_{\alpha}z}+ {\bf e}_y e^{-ik_{\alpha}z} \Big)\,, \;\;\alpha = 1,2 \,,
\end{equation} 
where $k_\alpha = \omega_\alpha/c$ are the wave vectors (c is the speed of light), and $E_{0\alpha}$ are the amplitudes of the corresponding running wave. We choose $\delta_1$ and $\delta_2$ as the detunings of the frequency components $\omega_1$ and $\omega_2$ from the resonance with the optical transition $F_g=1/2\to F_e=3/2$ (Fig.\,\ref{fig:F1}). In this case, both fields have close wave vectors $k_2 \simeq k_1 = k$, $\Delta k = k_2-k_1 \ll k$, and their relative phase $\Delta \phi = \Delta k\, z = (\delta_2-\delta_1) z/c$ is a slow function of the coordinate, which determines the macroscopic period of the bichromatic potential \cite{pru2013,pru2017,pru2023}
\begin{equation}\label{Lambb}
\Lambda = \frac{\pi c}{|\omega_2-\omega_1|}=\frac{\pi c}{|\delta_2-\delta_1|} \, .
\end{equation}
To form dissipative mechanisms of laser cooling, we choose detuning $\delta_2$ is close to resonance with the transition $F_g=1/2\to F_e=3/2$. Then, as follows from (\ref{Lambb}), the value $|\delta_1|$ must be chosen in the range $5-30$ GHz to form a centimeter-scale macroscopic potential.

Fig.\,\ref{fig:F3} shows a typical dependence of the macroscopic optical potential formed for $^{171}$Yb atoms in  double lin$\perp$lin configuration (\ref{double_linlin}). For example, for detunings $\delta_1 = -346 \gamma$ ($\simeq 10\,$GHz), $\delta_2 = -3 \gamma$, the period of macroscopic potential is $\Lambda = 1.2$ cm. In this case, for the intensities of the running  waves $I_1 \simeq 57$ W/cm$^2$ ($\Omega_{01} \simeq 22 \gamma$) and $I_2 \simeq 0.4$ W/cm$^2$ ($\Omega_{02} \simeq 1.9 \gamma$) the depth of the optical potential reaches $\Delta U_{opt} \simeq 6400\,\hbar \gamma$, (or $\sim 8.8\,$K in temperature units), which is comparable with the depth of magneto-optical traps and is quite sufficient for capturing, cooling and holding of atoms.

\begin{figure}[b]
\centering
\centerline{\includegraphics[width=3 in]{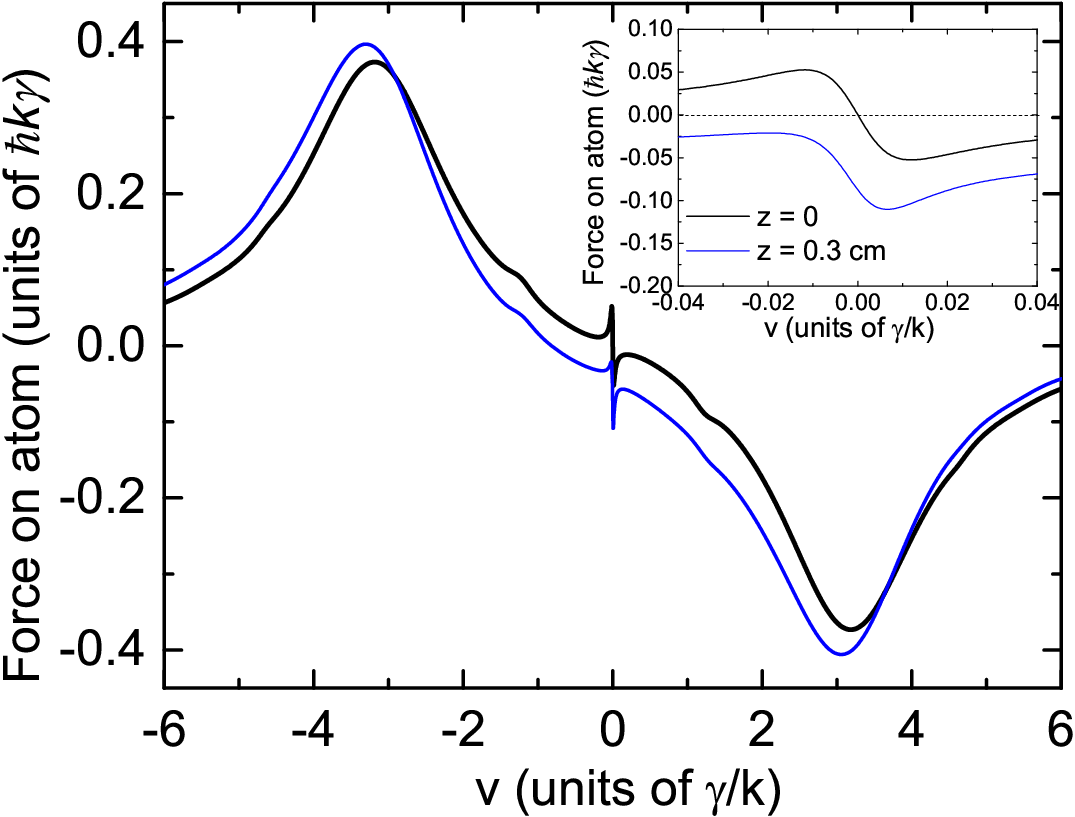}}
\caption{Force on an atom moving with velocity $v$ in a bichromatic field in the region of the global minimum of the optical potential $z=0$ (black line) and on the slope of the optical potential $z=0.3$ cm (blue line). The field parameters correspond to Fig.\ref{fig:F3}.}
\label{fig:F4}
\end{figure}

The dependences of the force on the velocity at the center and at the slope of the optical potential are shown in Fig.\,\ref{fig:F4}. As can be seen, in the low-velocity region the force has a typical dependence of sub-Doppler mechanisms of laser cooling \cite{dalibard89}. Also, for velocities $v = \pm |\delta_2|/k$ there are maxima caused by the resonance of atoms of a given velocity group with counter traveling waves of ${\bf E}_2$ field as a result of the Doppler effect. Note that the presence of another hyperfine component $F_{e_2}=1/2$ leads to additional resonances in the dependence of the force on the velocity of those spaced by $\delta v = \Delta_{hf}/k \simeq 11 \gamma/k$ (outside the region of Fig. 4). When shifting from the center of the optical trap,  the effect of rectification of the dipole force in a wide range of velocities is clearly seen (see the blue line in Fig.\ref{fig:F4}).
\begin{figure}[t]
    \centering
    \centerline{\includegraphics[width=3 in]{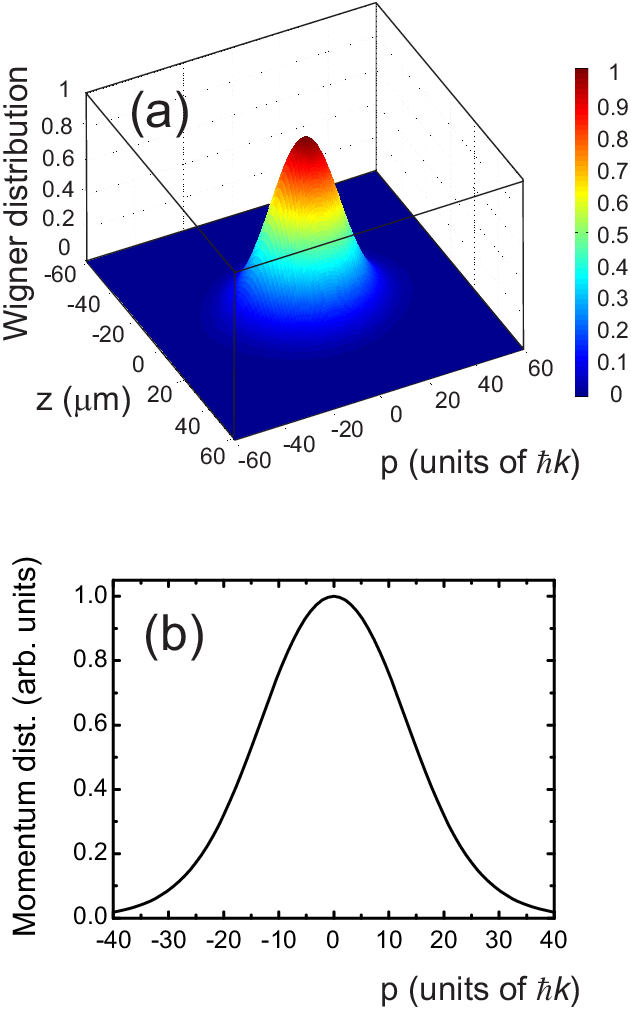}}
    \caption{The Wigner function of (a) the atomic phase distribution
${\cal F}(z, p)$ and (b) the momentum distribution of cold atoms in the macroscopic potential of the bichromatic field of double $lin \perp lin$ configuration. The field parameters correspond to Fig.\ref{fig:F3}.}
    \label{fig:F5}
\end{figure}

The temperature and size of trapped atoms in such a potential can be obtained from steady-state solution of the Fokker-Planck equation (\ref{FP}). For the parameters of Fig.\,\ref{fig:F3}, the laser cooling temperature reaches sub-Doppler values of $T \simeq 0.1\, \hbar \gamma/k_B \sim 130\,\mu$K (see Fig.\,\ref{fig:F5}), and the size of the atomic cloud is $\sim 28 \,\mu$m. Note that the Doppler temperature achievable in the MOT at the transition under consideration $F_g=1/2\to F_e=3/2$ is of the order of $T_D\simeq 0.5 \hbar \gamma/k_B \sim 700\,\mu$K, because  of sub-Doppler cooling mechanisms in standard MOT formed by field of the $\sigma_+-\sigma_-$ configuration are absent for this transition \cite{dalibard89}.

\begin{figure}
    \centering
    \centerline{\includegraphics[width=2.5 in]{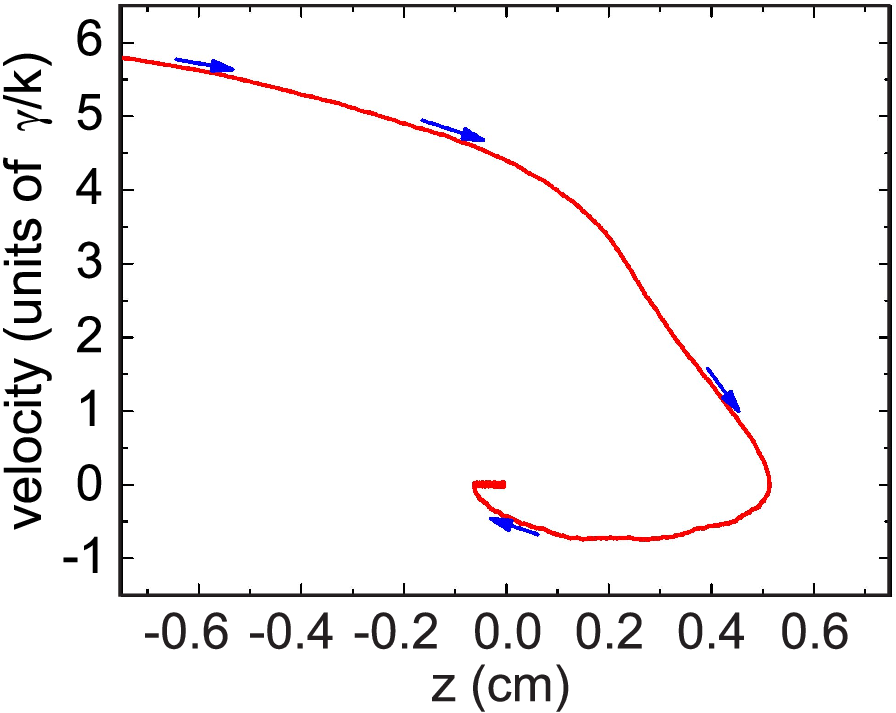}}
    \caption{The phase-space trajectory of an atom entering to the
macroscopic optical trap formed by double $lin \perp lin$ configuration of bichromatic field with parameters of Fig.\ref{fig:F3}. The
initial velocity of atoms on the trap boundary $v_{c} = 5.8 \gamma/k \simeq 67 $\,m/s. The 
blue arrows indicate the direction of the trajectory evolution.}
    \label{fig:F6}
\end{figure}

One of the important characteristics of a dissipative trap is the ability to capture atoms directly from the hot vapor. The number of captured atoms can be estimated from the relation presented in \cite{Monroe90,Gibble92}
\begin{equation}
N_c=\frac{L^2}{\sigma_c} \left( \frac{v_c}{u} \right)^4 \, ,
\end{equation}
where $v_c$ is the velocity at which atoms are captured in the trap, $u=(2k_BT/M)^{1/2}$ is the most probable velocity of atoms in a vapor ($170$ m/s for ytterbium atoms at $T=300$ K), $L$ is the size of the trap, and $\sigma_c$ is the cross-section for an atom to eject an atom from the trap. Therefore, the number of trapped atoms  can be increased via the use of field configurations that maximize $v_c$, or by using additional Zeeman slower of the atomic beam to effectively reduce $u$. The Fig.\,\ref{fig:F6} shows the trajectory of an atom in the phase space $(z,v)$ entering into the trap. The trajectory is calculated based on the Langevin equations (see, for example, \cite{Langevin,Langevin2}) describing the atom trajectory under the light force and associated stochastic diffusion.
The velocity at which atoms are able to be captured into the trap is $v_c \simeq 5.8 \gamma/k \simeq 67\,$ m/s for the parameters of Fig.\,\ref{fig:F3}. This leads to a theoretical estimate of the number of captured atoms $N_c \simeq 4 \times 10^{11}$, which is comparable to the number of atoms captured in a magneto-optical trap.

\begin{figure}
    \centering
    \centerline{\includegraphics[width=3.4 in]{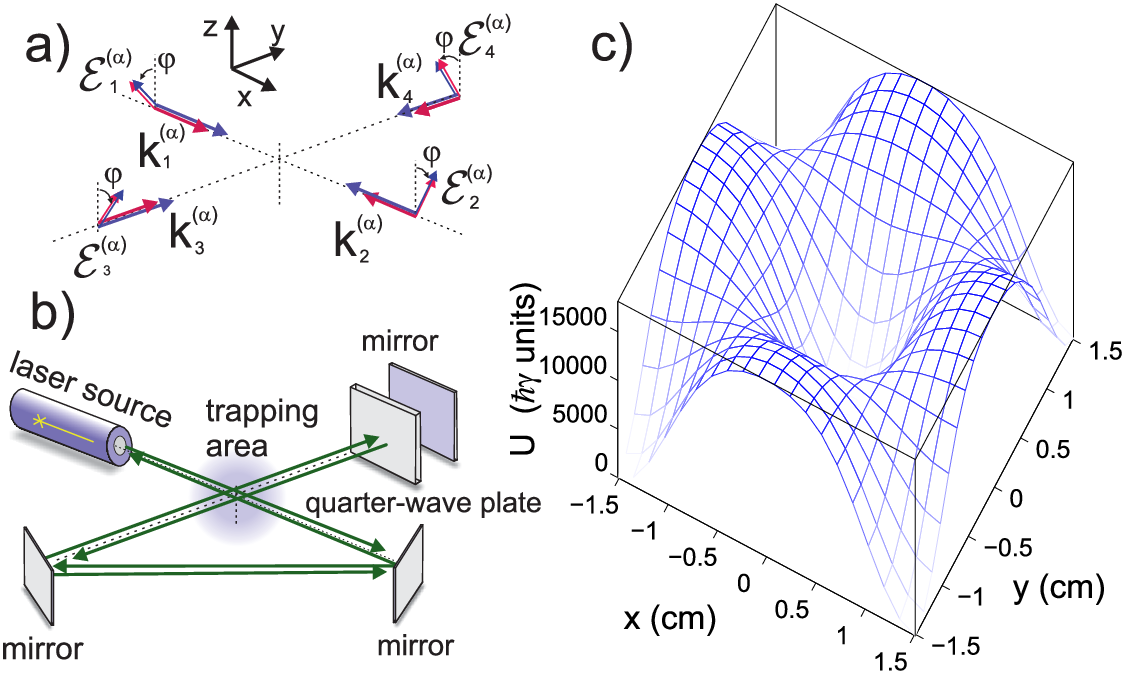}}
    \caption{Macroscopic 2D optical potential formed by four beams of bichromatic light waves with linear polarizations oriented at an angle of $\varphi =
\pi/4$ to the $z$ axis (a). The spatial configuration of the field is formed by the geometry of a folded standing wave (b). The macroscopic optical potential (c) has a trapping area depth is $\Delta U \simeq 15200\, \hbar \gamma \simeq 21$ K. The light field intensity and frequencies of the running waves forming the field  corresponds to parameters of Fig.\ref{fig:F3}. }
    \label{fig:F7}
\end{figure}

\section{Multidimensional  optical lattices of macroscopic scale}
2D and 3D configurations of macroscopic optical lattices can be formed in space by superimposing several traveling bichromatic waves. In this case, for a bichromatic lattice \cite{pru2023}, as well as for optical lattice formed by monochromatic light waves, the spatial topology can depend on the relative phase of the waves forming the field \cite{jessen96, Hemmerich1993}. Therefore, the optical lattices with phase-independent topology are of the greatest practical interest. Such lattices can be generally formed by a number of waves $M$ not exceeding $N+1$, where $N$ is the dimensionality of space. Thus, a 2D configuration of a phase-independent macroscopic optical lattice is formed by a maximum of three traveling bichromatic waves, and a 3D configuration –- by a maximum of four pairs \cite{pru2023}.

Alternatively, an optical lattice with phase-independent topology can be obtained by combining counter-propagating waves formed by a folded standing wave in both two-dimensional and three-dimensional configurations (Fig.\ref{fig:F7}). In such a field, the phases of the traveling waves are linked, and mirror displacement does not lead to a phase-induced change in the potential shape, but only to its spatial shift \cite{Domenico}. As can be seen (see Fig.\ref{fig:F7}) a 2D configuration with a folded standing wave geometry forms a macroscopic potential in the shape of a well for trapping and cooling atoms. Its depth agrees well with estimates obtained in the one-dimensional model (see Fig. \ref{fig:F3}). A 3D configuration of a macroscopic optical trap can be formed in a similar manner.

Thus, 2D and 3D optical traps, formed by $lin \perp lin$ configuration in each direction, are an alternative to standard magneto-optical traps for capturing atoms from the hot fraction and for laser cooling atoms to sub-Doppler temperatures.

\section{Conclusion}
In this paper we show that a deep purely optical dissipative macroscopic potential formed by a bichromatic field can be created for alkaline earth and similar atoms. Such a potential can capture and cool atoms directly from the hot fraction, providing an alternative to MOT.
Direct numerical calculations were performed using as the example of
$^{171}$Yb atoms in the field of a double $lin \perp lin$ configuration of counterpropagating waves. In such a field, it is possible to achieve sub-Doppler temperatures $T\simeq 130 \mu$K (see Fig.\,\ref{fig:F5}) of $^{171}$Yb  trapped atoms, which is unattainable for these atoms in the MOT. In addition, the size of the cold atoms cloud reaches submillimeter values, and the number of atoms is comparable to the number of atoms in the MOT. The presented purely optical trap can be considered as an alternative to the MOT for quantum sensors, quantum memory and optical frequency standards, which require precise control of the magnetic field in the region of cold atom interrogation.

In conclusion we note, that alkaline-earth and similar atoms possess only a nuclear magnetic moment in the ground state, which is three orders of magnitude smaller than the electronic one. This significantly reduces the sensitivity of proposed laser cooling scheme to residual magnetic fields. Our estimates show that the presence of a residual magnetic field $B < 1\,$G does not lead to substantial changes in the presented results. At the same time, such conditions are quite feasible and represent a standard requirement for experimental setups.

\begin{acknowledgments}
The research was supported by grant No. 25-22-00314 from the Russian Science Foundation, https://rscf.ru/project/25-22-00314/

\end{acknowledgments}

\section*{Data Availability Statement}
Data available on request from the authors

\appendix
\section{}

\setcounter{equation}{0}
\renewcommand{\theequation}{\thesection.\arabic{equation}}
The explicit form of the action of the operators $\hat{{\cal L}}^{(k)}$ on an matrix $\hat{\sigma}$ can be written in the form of the equations are given below. The zeroth order  $\hat{{\cal L}}^{(0)}$ of the expansion in the recoil parameter $-ikq_j$ has the form

\begin{eqnarray}\label{block0} 
\hat{{\cal L}}^{(0)}\{\hat{\sigma}\}^{ee} &=&
-\gamma\, \hat{\sigma}^{ee}
 - \frac{i}{\hbar} \Big[{\hat H}_0^{ee}{\hat \sigma}^{ee}-{\hat \sigma}^{ee}{\hat H}_0^{ee}\Big] \nonumber \\
&-&i\sum_{\alpha=1,2}\left( \hat{V}_\alpha({\bf r})
\hat{\sigma}_\alpha^{ge}-\hat{\sigma}_\alpha^{eg}\hat{V}^{\dagger}_\alpha({\bf r})
\right)\,,
\nonumber \\
\hat{{\cal L}}^{(0)}\{\hat{\sigma}\}^{gg} &=& \gamma\,\sum_{s=0,\pm1}
{\hat D}_s^{\dagger}\hat{\sigma}^{ee}{\hat D}_s - \frac{i}{\hbar} \Big[{\hat H}_0^{gg}{\hat \sigma}^{gg}-{\hat \sigma}^{gg}{\hat H}_0^{gg}\Big] \nonumber \\&-&i\sum_{\alpha=1,2}\left(\hat{V}_\alpha^{\dagger}({\bf r})
\hat{\sigma}_\alpha^{eg} -\hat{\sigma}_\alpha^{ge}\hat{V}_\alpha({\bf r})
\right)\,,  \\
\hat{{\cal L}}^{(0)}\{\hat{\sigma}\}^{eg}_\alpha &=&
-\left(\gamma/2-i\omega_{\alpha}\right)\hat{\sigma}^{eg}_\alpha  - \frac{i}{\hbar} \Big[{\hat H}_0^{ee}{\hat \sigma}^{eg}-{\hat \sigma}^{eg}{\hat H}_0^{gg}\Big] \nonumber \\
&-&i\left(
\hat{V}_\alpha({\bf r}) \hat{\sigma}^{gg}
-\hat{\sigma}^{ee}\hat{V}_\alpha({\bf r}) \right)\,,\;\; \alpha=1,2\,, \nonumber \\
\hat{{\cal L}}^{(0)}\{\hat{\sigma}\}^{ge}_\alpha &=&
-\left(\gamma/2+i\omega_\alpha\right)\hat{\sigma}^{ge}_\alpha   - \frac{i}{\hbar} \Big[{\hat H}_0^{gg}{\hat \sigma}^{ge}-{\hat \sigma}^{ge}{\hat H}_0^{ee}\Big]  \nonumber \\
&-&i\left(
\hat{V}_\alpha^{\dagger}({\bf r})\hat{\sigma}^{ee}
-\hat{\sigma}^{gg}\hat{V}_\alpha^{\dagger}({\bf r})  \right)\,,\;\; \alpha=1,2\,. \nonumber
\end{eqnarray}

For the $j$-th Cartesian component of the operator $\hat{{\cal L}}^{(1)}$ we have
\begin{eqnarray}\label{block1} 
\hat{{\cal L}}^{(1)}_{j}\{\hat{\sigma}\}^{ee}
&=&-\frac{1}{2}\sum_{\alpha=1,2} \left( \hat{F}_{\alpha_j} \hat{\sigma}_\alpha^{ge}  +\hat{\sigma}_\alpha^{eg}\hat{F}_{\alpha_j}^{\dagger}  \right) \,,\nonumber \\
\hat{{\cal L}}^{(1)}_j\{\hat{\sigma}\}^{gg}
&=&-\frac{1}{2}\sum_{\alpha=1,2}\left(\hat{F}_{\alpha_j}^{\dagger}
\hat{\sigma}_\alpha^{eg}+
\hat{\sigma}_\alpha^{ge}\hat{F}_{\alpha_j}   \right) \,,\nonumber \\
\hat{{\cal L}}^{(1)}_j\{\hat{\sigma}\}^{eg}_\alpha
&=&-\frac{1}{2}\left(\hat{F}_{\alpha_j}
\hat{\sigma}^{gg} +\hat{\sigma}^{ee}\hat{F}_{\alpha_j} \right),\;\;\alpha=1,2\,,\nonumber \\
\hat{{\cal L}}^{(1)}_j\{\hat{\sigma}\}^{ge}_\alpha
&=&-\frac{1}{2}\left(\hat{F}_{\alpha_j}^{\dagger} \hat{\sigma}^{ee}
+\hat{\sigma}^{gg}\hat{F}_{\alpha_j}^{\dagger} \right),\;\;\alpha=1,2\,,\nonumber  \\
\end{eqnarray}
where $\hat{F}_{\alpha_j} = -\partial_{{\bf r}_j} \,\hat{V}_\alpha({\bf r})$ is the $j$-th component of the force operator.

The second order terms are
\begin{eqnarray}\label{block2}
\hat{{\cal L}}^{(2)}_{jj'}\{\hat{\sigma}\}^{ee}
&=&\frac{i}{8}\sum_{\alpha=1,2}\left[\big(\partial_{{\bf r}_j} \partial_{{\bf r}_j'}\hat{V}_\alpha \big)
\hat{\sigma}_{\alpha}^{ge} -
\hat{\sigma}_{\alpha}^{eg}\big(\partial_{{\bf r}_j} \partial_{{\bf r}_j'}\hat{V}_{\alpha}^{\dagger}\big) \right] \,,\nonumber \\
\hat{{\cal L}}^{(2)}_{jj'}\{\hat{\sigma}\}^{gg}
&=&\frac{i}{8}\sum_{\alpha=1,2}\left[\big(\partial_{{\bf r}_j} \partial_{{\bf r}_j'}\hat{V}_{\alpha}^{\dagger}\big)
\hat{\sigma}_\alpha^{eg} - \hat{\sigma}_{\alpha}^{ge}\big(\partial_{{\bf r}_j} \partial_{{\bf r}_j'}\hat{V}_\alpha \big) \right] \nonumber \\
&+&{\hat \Gamma}^{(2)}_{jj'}\left\{{\hat \sigma}^{ee}\right\}\,, 
\\
\hat{{\cal L}}^{(2)}_{jj'}\{\hat{\sigma}\}^{eg}_{\alpha}
&=&\frac{i}{8}\left[\big(\partial_{{\bf r}_j} \partial_{{\bf r}_j'}\hat{V}_\alpha \big) \hat{\sigma}^{gg}
-\hat{\sigma}^{ee}\big(\partial_{{\bf r}_j} \partial_{{\bf r}_j'}\hat{V}_\alpha \big)
\right],\nonumber \\
\hat{{\cal L}}^{(2)}_{jj'}\{\hat{\sigma}\}^{ge}_\alpha
&=&\frac{i}{8}\left[\big(\partial_{{\bf r}_j} \partial_{{\bf r}_j'}\hat{V}_{\alpha}^{\dagger}\big)
\hat{\sigma}^{ee}
-\hat{\sigma}^{gg}\big(\partial_{{\bf r}_j} \partial_{{\bf r}_j'}\hat{V}_{\alpha}^{\dagger}\big) \right], \nonumber \\
\alpha&=&1,2\,, \nonumber
\end{eqnarray}
which contain contributions proportional to the second
derivative of the dipole interaction operators $\hat{V}_{\alpha}$.
Here ${\hat \Gamma}^{(2)}_{jj'}\left\{{\hat \sigma} \right\}$ is the second order of the expansion of the spontaneous relaxation operator (\ref{e:GE1}) in the recoil parameter
\begin{equation}
{\hat \Gamma}^{(2)}_{ij}\left\{{\hat \sigma} \right\} = \frac{\gamma}{5}\sum_{m,n}\Big( \delta_{ij}\delta_{mn}-\frac{1}{4}\big[\delta_{in}\delta_{jm}+\delta_{im}\delta_{jn}\big]\Big) {\hat D}^{\dagger}_m {\hat \sigma} {\hat D}_n \,.
\end{equation}
The presented operators $\hat{{\cal L}}^{(\kappa)}, (\kappa = 0,1,2)$ allow us to obtain expressions for the force and diffusion tensors on atom in a bichromatic field and to analyze the kinetics of atoms within the framework of the semiclassical Fokker-Planck equation (\ref{FP}).

\nocite{*}
\bibliography{main_text}

\end{document}